\documentclass[twocolumn,aps,pre,showpacs,preprintnumbers]{revtex4}
\usepackage{graphics}
\usepackage{epsfig} 
\usepackage{hyperref}
\usepackage{graphics}
\usepackage{color}      % use if color is used in text
\usepackage{graphicx}
\usepackage{graphicx}% Include figure files
\usepackage{dcolumn}% Align table columns on decimal point
\begin{document}
\title{The  effect of temperature on viscous friction and
the performance of  a  Brownian heat engine }
\author{Solomon Fekade Duki}
\affiliation {National Center for Biotechnology Information,
National Library of Medicine and National Institute of Health,
8600 Rockville Pike, Bethesda MD, 20894 USA}
\author{Mesfin Asfaw Taye}
\affiliation{Department of Physics and Astronomy, California
State University  Dominguez Hills, California, USA }
\date{\today}
\begin{abstract} 

We explore the transport features of a Brownian particle that walking in
a periodic ratchet potential that is coupled with a spatially varying
temperature background. Since the viscous friction of the medium
decreases as the temperature of the medium increases, any reasonable
exploration regarding the thermodynamic features of the Brownian
engine should take into account the role of temperature on the
viscosity of the fluid. In this work, we study this effect of
temperature by considering a viscous friction that decreases
exponentially as the background temperature increases. Our result
depicts that the Brownian particle exhibits a fast unidirectional
motion when the viscous friction is temperature dependent than
that of constant viscous friction. Moreover the efficiency of
this motor is considerably enhanced  when the viscous friction
is temperature dependent. On the hand, the motor exhibits a higher
performance of the refrigerator  when the viscose friction is
taken to be constant.
\end{abstract}
\pacs{Valid PACS appear here}% PACS, the Physics and Astronomy
\maketitle
 
\section {Introduction}
Over the past decade or two there has been a great interest in the study of noise-induced
transport features of micron and nanometer sized particles. This was motivated not only
for a better understanding of the nonequilibrium statistical physics of such systems 
but also due to the desire to construct artificial tiny motors that operate at the microscopic
or nanoscopic levels \cite{c1,c2}. Several studies have shown that the dynamics of
these particles exhibit a unidirectional motion when they are exposed to a bistable potential
that is subjected to a spatial or temporal symmetry breaking fields such as inhomogeneous
temperature background \cite{cc10,cc11,cc12,c3,c4,c5,c6,c7}. In particular when the
bistable potential is exposed to a spatially varying temperature,  the particles will have
a fast unidirectional motion where the intensity of the current rectification depends
strongly on the strength of background temperature and the potential barrier height.
While undergoing a rectified motion along the reaction coordinate, these particles
perform a useful work at the expense of the heat taken out from the hotter reservoir
thus act as a Brownian heat engines. For detailed practical applications as
well as the characteristics, and working principle of these classical Brownian
motors, the reader is encouraged to  refer to the works of Peter H\"anggi and his collaborators \cite{c8,cc8}.

Several studies have been also conducted to understand the factors that affect performance of
a Brownian engine that is driven by a spatially varying temperature. These studies have
explored the different operational regimes of the engine both at the quasistatic limit
and when the engines operate at finite time interval \cite{c9,c10,c11,c12,c13,c14,c15,c16}.
In line with these studies, we have explored in our previous work \cite{c17}
the effect of thermal inhomogeneity on the performance of such heat engines
by considering a Brownian particle in a ratchet potential that moves through
a highly viscous medium. In the model considered in \cite{c17}, the Brownian
particle is driven by the thermal kick it receives from a linearly decreasing
background temperature. The study showed that even though the energy transfer
due to kinetic energy is neglected, Carnot efficiency cannot be achieved at
quasistatic limit. At quasistatic limit, the efficiency for such a Brownian
heat engine approaches the efficiency ($\eta$) of an endoreversible heat
engine; {\em i. e.} $\eta= 1-\sqrt{{T_{h}\over T_{c}}}$ \cite{c18}. More
recently, by considering a Brownian motor that operates  between two
different heat baths, we have also explored both the nonequilibrium steady state (NESS) and short time behavior of the engine \cite{c19}. This
investigation studied the thermodynamic feature of the engine for both
the isothermal case with a load and  nonisothermal case with
and without a load.

So far most of the studies of Brownian heat engines considered only temperature
invariance viscous friction. However, it is well know that the viscosities of
different media tend to depend on the intensity of the background temperature
\cite{Landau}. In liquid or glassy meadium viscosity tend to decrease when the
intensity of the background temperature increases.  
This is due to the fact that an increase in temperature of the
medium brings more agitation to the molecules in the medium, and hence increases
their speed. This speedy motion of the molecules creates a reduction in interaction
time between neighboring  molecules. In turn, at macroscopic level, there will be
a reduction in the intermolecular force, and hence reduced viscosity of the fluid.
Consequently, as the temperature of the viscous medium decreases, the viscous
friction in the medium decreases. Meanwhile for a position dependent temperature
along the reaction coordinate, the viscous friction is also spatially dependent.
In such case we want to stress that the effect of temperature on the particles’
mobility and performance of the motor will be twofold. First, it directly assists
the particles to surmount the potential barrier; {\em i. e.} particles jump the
potential barrier at the expenses of the thermal kicks. Second, when temperature
increases, the viscous friction gets attenuated and particles experiences a reduced
inertial effect, which in turn increases the particles mobility and efficiency.

In this paper, we address the role of temperature dependent viscous friction on the Brownian
heat engine by considering an exponential temperature dependence of friction,
$\gamma(x)=Be^{-A T(x)}$, as  proposed originally by Reynolds \cite{c20}.
Our analysis  shows  that whether $\gamma$ is temperature dependent or not,  at
quasistatic limit one always gets a Carnot efficiency and a Carnot refrigerator
so long as the heat exchange via kinetic energy is omitted. However, when the heat
exchange via the kinetic energy is included, it will be impossible to attain the 
Carnot efficiency or Carnot refrigerator even at quasistatic limit. Moreover, far
from quasistatic limit, the engine exhibits an enhanced  performance when the
viscose friction is taken to be temperature dependent.

The rest of the paper is organized as follows. In section II, we present our model
for the system. In section III, by considering a viscous friction that decreases
exponentially with temperature, we explore the dependence of mobility, efficiency
and performance of the refrigerator on the the model parameters. Finally we give
summary and conclusion of the paper in Section IV.

\section{The model}  
We consider a Brownian particle that rattles in a one dimensional piecewise linear bistable
potential with an external load; {\em i. e.} $U(x)=U_{s}(x) + fx$, where the ratchet potential
$U_{s}(x)$ is described by
\begin{equation}
  U_{s}(x)=\left\{
  \begin{array}{ll}
    2U_{0}\left({x\over L_{0}}\right),& \text{if}~~~ 0 \le x \le {L_{0}\over 2};\\ %cr
    2U_{0}\left(1-{x\over L_{0}}\right),& \text{if} ~~~{L_{0}\over 2} \le x \le L_{0}. %\cr
   \end{array}
   \right.
\end{equation}
Here $U_{0}$ and $L_{0}$ denote  the barrier height and the width of the ratchet potential,
respectively, and $f$ is the strength of the load. The potential exhibit its maximum value
$U_{0}$ at  $x={L_{0}\over 2}$ and its minima at $x=0$  and $x={L_{0}}$.
The background temperature in the system is taken to be spatially varying hot and cold regions
where
\begin{equation}
T(x)=\left\{
\begin{array}{ll}
T_{h},& \text{if} ~~~0 \le x \le {L_{0}\over 2};\\
T_{c},& \text{if} ~~~ {L_{0}\over 2} \le x \le L_{0}
\end{array}
\right.
\end{equation}
 as shown in Fig. 1.  In this model both the potential $U_{s}(x)$ and the
 background temperature $T(x)$ are assumed to be periodic with period $L_0$;
 {\em i. e.} $U_{s}(x+L_{0})=U_{s}(x)$ and $T(x+L_{0})=T(x)$.
\begin{figure}[ht]
\centering
%\subfigure[Bild a.] % caption for subfigure a
{
    %\label{fig:sub:a}
  \includegraphics[width=8cm]{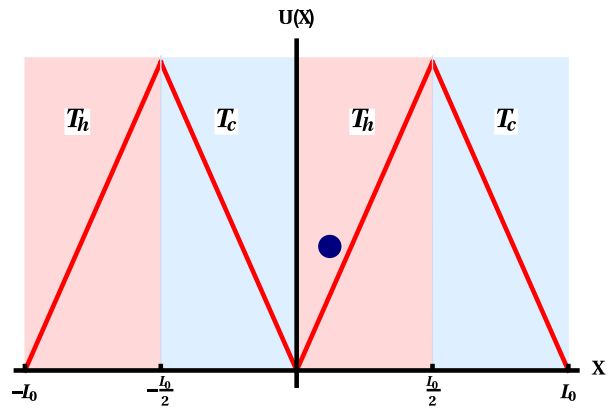}
}
\caption{(Color online) Schematic diagram for a particle  in a piecewise linear bistable potential
in the absence of an external load. The potential exhibits a potential maxima $U_{0}$ at
$x={L_{0}\over 2}$. The  potential minima is located at $x=0$  and $x=L_{0}$.   Due to the
thermal background kicks, the particle ultimately attains a steady state velocity as long
as a  temperature difference between the hot and cold reservoirs is retained.}
\label{fig:ratchet}
\end{figure}

For a Brownian particle that is arranged to undergo a random walk in a highly
viscous medium, regardless of the magnitude of the external bias force, the
particle has very little chance to accelerate along the medium. Hence one can
safely neglect the inertial contribution to the Langevin equation and the
dynamics of the particle under the influence of external potential $U(x)$ is
governed  by
 \begin{eqnarray}
\gamma(x){dx\over dt}&=&-{\partial U(x)\over \partial x} +
\sqrt{2k_{B}\gamma(x) T(x)}\xi(t),
\end{eqnarray}
where $\gamma$ is the viscous friction, and $k_B$ is the Boltzmann's constant.
As mentioned earlier, in this work the viscous friction is taken to have an
exponential temperature dependence of the form $\gamma(x)=Be^{-A T(x)}$
where $A$ and $B$ are constants. Since $T(x)$ is spatially variable we have,
 \begin{equation}
 \gamma(x)=\left\{
  \begin{array}{ll}
    Be^{-A T_{h}},& \text{if}~~~ 0 \le x \le {L_{0}\over 2};\\
    Be^{-A T_{c}},&\text{if}~~~  {L_{0}\over 2} \le x \le L_{0}.
 \end{array}
\right.
\end{equation}
In this system the random noise $\xi(t)$ is assumed to be Gaussian and white, satisfying
$\left\langle  \xi(t) \right\rangle =0$ and $\left\langle \xi(t)  \xi(t') \right\rangle=\delta(t-t')$.
Hereafter we will work on the units where $k_{B}$ and $B$ are unity.

In the high friction limit, the dynamics of the Brownian particle is governed by \cite{c9}
\begin{equation}
{\partial P(x,t)\over \partial t}={\partial\over  \partial x}
\left[{{\gamma}^{-1} }\left(
U'(x)P(x,t)+{\partial \over \partial x}\left(T(x)P(x,t)\right)\right)\right]
\end{equation}
where $P(x,t)$ is the probability density of finding the particle at position $x$ at
and at time $t$, and $U'(x)={d\over dx}U$.  At steady state the current is given by
$J(x) = -\left[U'(x)P_{s}(x) + {\partial \over \partial x}\left[{T(x)}P_{s}(x)\right]\right]$.
For periodic boundary condition, $P_s(x +L_{0}) = P_s(x)$, the corresponding steady state current $J(x)$
can be evaluated exactly using the same approach shown in \cite{c9}. After some algebra,
the closed form expression for the steady state current is given as 
\begin{equation}
J= -{\varsigma_{1}\over \varsigma_{2}\varsigma_{3}+(\varsigma_{4}+\varsigma_{5})\varsigma_{1}}
\end{equation}
where the expressions  for $\varsigma_{1}$, $\varsigma_{2}$, $\varsigma_{3}$, and $\varsigma_{4}$
are given by
\begin{eqnarray}
\varsigma_{1}& = &-1+e^{{L_0(f-{2U_{0}\over L_0})\over 2T_{c}}+{L_0(f+{2U_{0}\over L_0})\over 2T_{h}}}\\ \nonumber
\varsigma_{2}& = &{e^{-{{fL_{0}(T_{c}+T_{h})\over T_{c}}+2U_{0}\over 2T_{h}}}\left(e^{fL_{0}\over
                    2T_{c}}-e^{U_{0}\over T_{c}}\right)L_0\over fL_{0}-2U_{0}}\\ \nonumber
          &&-{\left(e^{-{fL_{0}+2U_{0}\over 2T_{h}}}-1\right)L_{0}\over fL_{0}+2U_{0}} \\ \nonumber
\varsigma_{3} & = &{e^{-AT_{c}-{U_{0}\over T_{c}}+{fL_{0}+2U_{0}\over 2T_{h}}}
\left(e^{fL_{0}\over 2T_{c}}-e^{U_{0}\over T_{c}}\right)L_0T_{c}\over fL_{0}-2U_{0}}+\\ \nonumber
            &&{e^{-AT_{h}}\left(e^{fL_{0}+2U_{0}\over 2T_{h}}-1\right)L_{0}T_{h}\over fL_{0}+2U_{0}}\\ \nonumber
\varsigma_{4}& = &{e^{-AT_{h}}L_{0}^2\left(fL_{0}+2((-1+e^{-{fL_{0}+2U_{0}\over 2T_{h}}})T_{h}+U_{0})\over
                  2(fL_{0}+2U_{0})^2\right)}
\end{eqnarray}
and $\varsigma_{5}=L_{0}^2(t_1+t_2+t_3+t_4)$. Here $t_1$, $t_2$, $t_3$, and $t_4$ are given by
  \begin{eqnarray}
  t_1&=&{e^{-AT_{c}}\over 2fL_{0}-4U_{0}}\\ \nonumber
  t_2&=&{e^{-AT_{c}}(1-e^{-{{fL_{0}-2U_0}\over 2T_{c}}})T_c\over (fL_{0}-2U_{0})^2}\\ \nonumber
  t_{3}&=&{e^{-AT_{h}}(1-e^{-{{fL_{0}-2U_0}\over 2T_{c}}})T_h\over (f^2L_{0}^2-4U_{0}^2)} \\ \nonumber
  t_{4}&=&{e^{-AT_{h}-{fL_{0}(T_{c}+T_{h})\over 2T_{c}T_{h}}-{U_{0}\over T_{h}}}(-e^{{fL_{0}\over
  2T_{c}}}+e^{{U_0\over T_{c}}})T_{h}\over (f^2L_{0}^2-4U_{0}^2)}.
  \end{eqnarray}
  
In the absence of external load, {\em i. e.} $f=0$, Eq. (6) reduces to
\begin{eqnarray}
J= { {4(z_1z_2)U_0^2} \over {
L_0^2 (\psi_1 + \psi_2 + {
{ 2 (z_1T_c + z_2 T_h) (1 + e^{{U_0} \over {T_c}}(-2 + e^{ {U_0} \over {T_c}})}
\over { e^{ {U_0} \over {T_c} } - e^{ {U_0} \over {T_h} } } }
}
}
\end{eqnarray}
where $\psi_1 = -2z_1T_c - 2z_2T_h$, $\psi_2 = -z_2U_0 + z_1U_0$, $z_1 = e^{AT_h}$
and $z_2 = e^{AT_c}$.

When $A=0$, the model will be reduced to a constant viscous friction $\gamma$,
the case that was studied before and Eq. (9) reproduces the result of \cite{c9},
\begin{equation}
J^C={2U_{0}^2\over (T_h+T_c)}\left[{1\over e^{{U_{0}\over T_h}}-1}-{1\over e^{{U_{0}\over T_c}}-1}\right].
  \end{equation}
For a small barrier height (small  $U_{0}$), the steady state current converges to
\begin{equation}
J\approx{4U_{0}(T_{h}-T_{c})e^{A(T_{h}+T_{c})}\over L_{0}^2 (T_{h}+T_{c})(e^{AT_{c}}+e^{AT_{h}})}.
\end{equation}
On the other hand for large $U_{0}$  the current can be approximated as
    \begin{equation}
J\approx{2U_{0}^2e^{A(T_{h}+T_{c})-{U_{0}\over T_{h}}}\over L_{0}^2 (e^{AT_{c}}T_{h}+e^{AT_{h}}T_{c})}.
\end{equation}
The drift velocity $V$ of the Brownian particle is associated to the steady state current $J$ via 
 $V=L_{0}J$.

As discussed before, a non-vanishing particle current $J$ can be obtained as long
as a distinct temperature difference between the hot and cold reservoirs is maintained
even in the absence of a load. However, the external load is necessary to get a non-vanishing
current for isothermal  symmetric ratchet. Whether isothermal or not, the direction
of the current is always dictated by the load. In the regime
where $J>0$, the model acts as a heat engine and in this case, in one cycle, a minimum
$(U_{0}+\gamma^* J{L_{0}^2\over 4}+f{L_{0}\over 2})$ energy per particle
is needed to overcome the viscous drag force $\gamma^* V/2$, the potential barrier
$U_{0}$  and the external load $f$. Here $\gamma^*=B(e^{-AT_{h}}+e^{-AT_{c}})$.
In addition, an amount of ${1\over 2} { k_B(T_h-T_c)}$  energy per cycle is
transferred from the hot to the cold heat bath via the kinetic energy at the boundaries
of the heat baths. Thus for arbitrary particle crossing through the potential barrier,
the amount of heat energy taken from the hot reservoir in a given cycle  is given by
 \begin{equation}
Q_{h}=\left(U_{0}+\gamma^* J{L_{0}^2\over 4}+f{L_{0}\over 2}+{1\over 2}k_{B}(T_{h}-T_{c})\right).
\end{equation}
On the other hand the heat given to the cold reservoir  is 
 \begin{equation}
Q_{c}=\left(U_{0}-\gamma^* J{L_{0}^2\over 4}-f{L_{0}\over 2}+{1\over 2}k_{B}(T_{h}-T_{c})\right).
\end{equation}
If the motor acts as a refrigerator,  the net heat flow  to the cold heat bath has a
magnitude \cite{c10}
\begin{equation}
Q_{c}=\left(U_{0}-\gamma^* J{L_{0}^2\over 4}-f{L_{0}\over 2}-{1\over 2}k_{B}(T_{h}-T_{c})\right).
\end{equation}

In one cycle, the particle does a work of $W=Q_h-Q_c$ against the load and the
viscous friction. Furthermore, the efficiency is given by $\eta= W/Q_{h}$.
The performance of the refrigerator is also given as $P_{ref}=Q_{c}/W^L$
where $W^L=fL_0$ is the work done by the load. Here it is also worth mentioning
that some motors are not designed to pull loads and in that case an
alternative measure for their efficiency depends on the task that each
motor performs. For example, some engines  may have
to achieve high velocity against a frictional drag. This practically
suggests that the motor could objectively used to move things a certain
distance in a given interval of time. In such motors (where $f=0$), the useful
work is calculated as $W=(Q_{h}-Q_{c})=\gamma^* J{L_{0}^2\over 2}$.
Once again we want to emphasize that the case where $A=0$ corresponds
to constant  $\gamma$. Otherwise when $A>0$, $\gamma$ is temperature
dependent and in this case we measure work in the unit where $A$ is taken
to be unity for convenience.

We now introduce the dimensionless quantities for energy, length, and time to
simplify the model equations. We measure energy in units of $k_BT_c$ (with $K_B=1$),
hence the load, temperature and barrier height are resacled as ${\bar f}=fL_{0}/T_{c}$,
${\bar T}(x)=T (x)/T_{c}$, and ${\bar U_{0}}=U_{0}/T_{c}$ respectively. Similarly
we rescale length and time as ${\bar x}=x/L_{0}$ and ${\bar t}=t/ \beta$ respectively.
Here $\beta=\gamma(x) L_{0}^2/T_{c}$ is the relaxation time. For convenience we
use $\tau=T_{h}/T_{c}$ to measure the temperature of the hot region.
From now on all equations will be expressed in terms of the dimensionless parameters
and, hence for brevity we dropped all the bars. We also work in a unit where $A=1$
for the non-constant viscous friction.

\section{ The role of viscosity}
Previous studies on Brownian heat engine working due to specially arranged  thermal
gradient  have given us an insight on how the engine thermodynamic features depend
on the model parameters.  These investigations depicted that  the particle attains
a unidirectional motions as long as a distinct temperature difference is retained
along the bistable potential.  In the presence of a load, the engine exhibits an
intriguing dynamics where  the magnitude of the load dictates  the  direction of
the particle flow. So far most of the studies assumed the viscosity of the medium
to be temperature independent. However, the viscosity of the medium indeed
significantly relies on the intensity of the background temperature along the
reaction coordinate.
\begin{figure}[ht!]
\centering
%\subfigure[Bild a.] % caption for subfigure a
{
    %\label{fig:sub:a}
  \includegraphics[width=8cm]{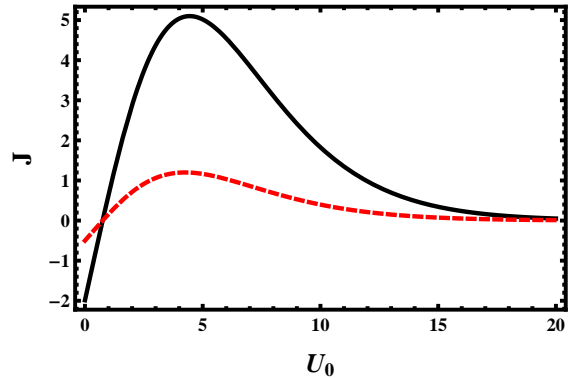}
}
\caption{The current $J$  as a function of $U_{0}$ for the parameter values of $f=0.3$,
and $\tau=2.0$. The black solid line stands for temperature dependent viscous
friction case while the red line exhibits the current for constant $\gamma$.  }
\end{figure}
\begin{figure}[ht!]
\centering
%\subfigure[Bild a.] % caption for subfigure a
{
  %\label{fig:sub:a}
  \includegraphics[width=8cm]{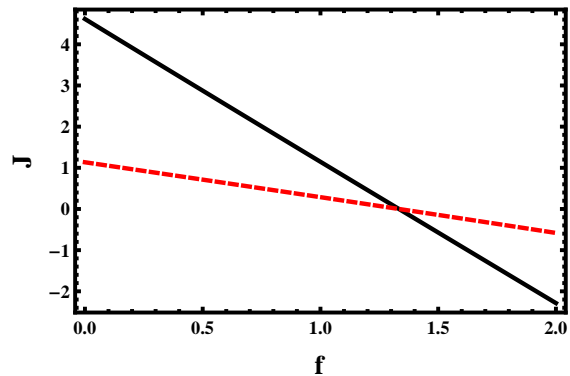}
}
\caption{The current  $J$ as a function of $f$ for parameter choice $\tau=2.0$ and
$U_{0}=2$. The black solid stands for the current that is evaluated  by taking $\gamma$
to be temperature dependent ($A=1$) while the red line is plotted by taking $\gamma$
to be constant ($A=0$).}
\end{figure}

In this section we will explore the role of temperature on the performance of the
 Brownian motor we considered. Here the viscous medium is assumed to vary
exponentially with the temperature. As shown in Eq. (6) the steady state current
(or equivalently the velocity) is solved exactly. Exploiting Eq. (6), one can see
that the mobility of the particles strictly relies on the potential barrier height.
When barrier height is too small, the particle moves sluggishly in the medium. In
the limit $ U_{0} \to 0$, $J \to 0$  depicting that in order to rectify the random
Brownian into a directed  motion, the presence of a bistable potential is vital.
In the high barrier limit, the particle again moves very slowly. Particularly,
when $U_{0} \to \infty$, the current $J$ goes to zero as the particle encounters
a difficulty in jumping the potential barrier height. To explore the dependence of
$J$ on $U_0$, we have shown in Fig. 2 the plot of $J$ as a function of $U_{0}$
for fixed load of $f=0.3$ and  $\tau=2.0$. The black solid
curve in the figure stands for the current evaluated for temperature dependent $\gamma$
(with $A=1$) while the red curve shows the constant $\gamma$ case where $A=0$. The figure
clearly indicates that the velocity of the
particle is significantly enhanced when the viscous friction of the medium is temperature
dependent.
\begin{figure}[ht!]
\centering
%\subfigure[Bild a.] % caption for subfigure a
{
\includegraphics[width=8cm]{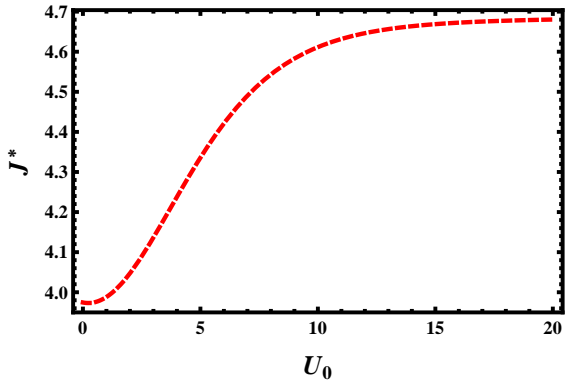}}
\caption{ The ratio of current between temperature dependent $\gamma$ ($A=1$) and
constant $\gamma$ ($A=0$) case as a function of $U_{0}$ for the  parameter values
of $f=0.3$ and   $\tau=2.0$. }
\end{figure}
\begin{figure}[ht]
\centering
{
%\label{fig:sub:a}
\includegraphics[width=8cm]{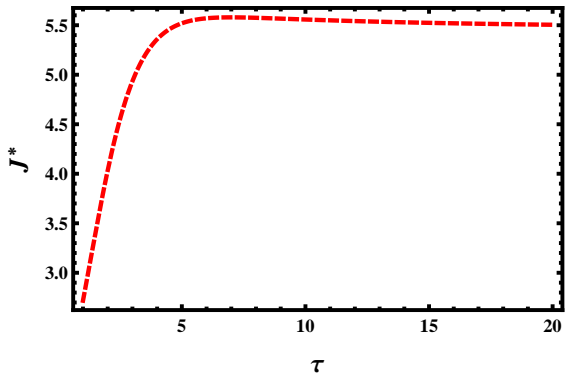}}
\caption{ The ratio of current between temperature dependent $\gamma$ ($A=1$) and
constant $\gamma$ ($A=0$) case   as a function of $\tau$ for the  parameter values
of $f=0.3$, and $U_{0}=2.0$.}
\end{figure}

In Fig. 3 the current $J$ is plotted as a function of the load $f$ for both
temperature dependent  and constant $\gamma$ cases. Once again the particle exhibits a
higher forward or backward velocity when $\gamma$ is temperature dependent. The
stall force, at which  the particle attains a zero velocity, is insensitive to
whether $\gamma$ is temperature dependent or not. Moreover, the direction of
particle velocity is dictated  by the magnitude of the load. A smaller load lacks
the capacity of reverse the current. However,  a larger load can reverse the particle
motion  since it has the capacity of renormalizing  the effect of temperature.

The ratio of the two current, $J^*=J/J^c$, between the temperature dependent $\gamma$
($A=1$) and constant $\gamma$ ($A=0$) cases, as a function of $U_{0}$ can be calculated
via Eq. (6).  In Fig. 4 $J^*$ is shown as a function $U_{0}$ for the  parameter
values of $f=0.3$ and   $\tau=2.0$. The figure depicts  that indeed the velocity
for temperature dependent  $\gamma$ case  is higher than that of constant $\gamma$.
Particularly, as the potential barrier increases, $J^* $ increases considerably.
This can be further appreciated by  explicitly evaluating $J^*$ in the limiting
cases. For $f=0$ case, Eq. (9) reduces to
\begin{equation}
J^*={{2 e^{1+\tau}(1+\tau)} \over
{
  {(-e + e^\tau) \over {-1 + e^{ {U_0}\over \tau}}} +
  {{2 e^{\tau+U_0} + 2 e^{1+U_0}\tau - e^{\tau}( 2 + U_0) + e(-2\tau + U_0) }
  \over {-1+e^{U_0}}}
  }
    }
\end{equation}
For small barrier height, Eq. (16) converges to
   \begin{equation}
J^*\approx{e^{(\tau +1)}\over (e+e^{\tau})}.
\end{equation}
$J^*>1$ showing that the mobility of the particle is high for $A=1$ (temperature
dependent $\gamma$ case). On the other hand for large $U_{0}$, $J^*$ is approximated
as
\begin{equation}
J^*\approx{e^{(\tau+1)}\over (e\tau+e^{\tau})}.
\end{equation}
Once again $J^*>1$ revealing  that the velocity of the particle is higher for $A=1$
(temperature dependent $\gamma$ case).

The plot of $J^* $  as a function of $\tau$ depicts that $J^* $  increases with
$\tau$ and attains an optimum value for a certain value of  $\tau$ (see Fig.5). It then decreases
when $\tau$ further increases. For  very small $\tau$, Eq. (16) leads to
$J^*={1\over 2}e(\tau+1)$. In the limit $\tau \to 1$, $J^* \to e$.

\begin{figure}[ht]
\centering
{
\includegraphics[width=8cm]{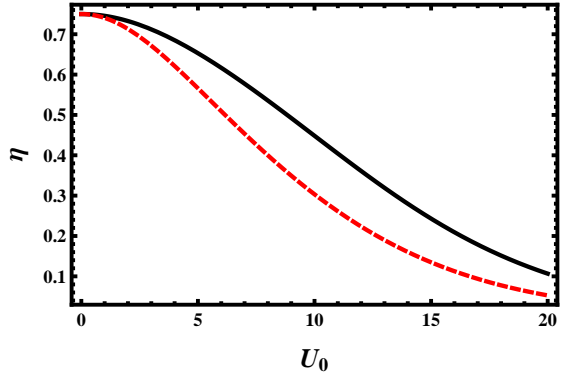}}
\caption{ The efficiency $\eta$ as a function of  $U_{0}$ for the  parameter v
values of $f=0.0$, and   $\tau=4.0$. }
\end{figure}
All these results indicate that the thermal background temperature undoubtedly
affects the strength of the viscosity of the medium and hence this effect
cannot be avoided. Particularly an enhanced mobility of the particle is
observed when the difference between the hot and cold regions in the
background temperatures is big. A similar effect is also observed 
when the potential barrier is increased.

We now explore the dependence of the efficiency $\eta$  and
the performance of the refrigerator $P_{ref}$  on the model parameters.
To start with we first look at how the performance of the engine
depends on the barrier height and the load by omitting the heat exchange
via kinetic energy. Fig. 6 shows the efficiency $\eta$ as a function
of $U_{0}$ for the  parameter values of $f=0.0$, and $\tau=4.0$. In this
figure, the solid line and dashed lines indicate the $A=1$ and $A=0$
cases respectively. For both cases, the $\eta$ decreases from its maximum
quasistatic efficiency  (Carnot efficiency) when $U_{0}$ increases. Far
from the quasistatic limit, the temperature dependent medium gives a
larger $\eta$ as compared to the constant $\gamma$ case.

It is important here to note that the quasistatic limit corresponds to the
case where $J \to 0$. The current approaches zero when $U_{0}\to 0$ for
zero external load.  In the presence external load, the particle
current vanishes when
\begin{equation}
f=2U_0{\left({\tau-1}\over {\tau+1}\right)}.
\end{equation}
We find that at quasistatic limit (for  both $A=1$ and $A=0$ cases),
the efficiency always goes to Carnot efficiency, {\em i. e.}
\begin{equation}
\eta=1-{1\over \tau}.
\end {equation}

\begin{figure}[ht]
\centering
%\subfigure[Bild a.] % caption for subfigure a
{
    %\label{fig:sub:a}
    \includegraphics[width=8cm]{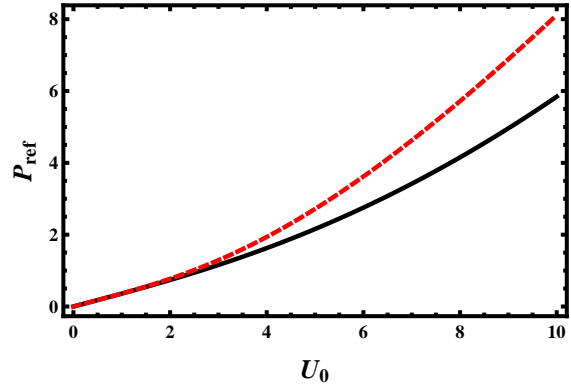}}
\caption{  $P_{ref}$ as a function of $U_0$ for the  parameter values of
$ f=0.8$, and   $\tau=6.0$. The solid black line stands the current which
is plotted  by taking  $\gamma$ to be temperature dependent while the red
dashed line is plotted by taking $\gamma$ to be constant. }
\end{figure}
  \begin{figure}[ht]
\centering
%\subfigure[Bild a.] % caption for subfigure a
{
    %\label{fig:sub:a}
    \includegraphics[width=8cm]{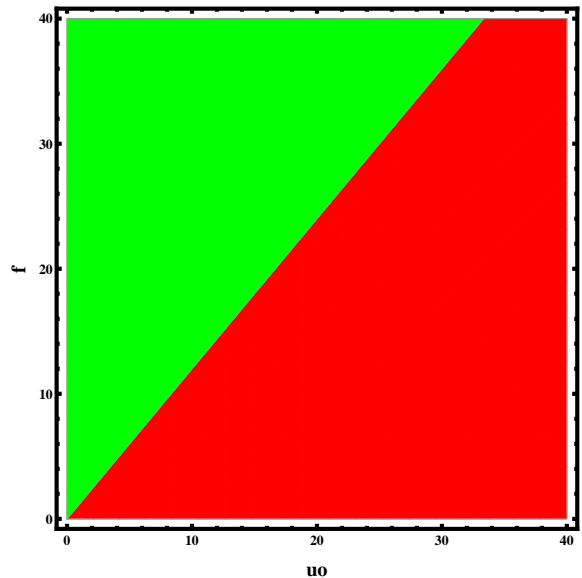}}
\caption{ The two operation region of   the engine in parameter space of
$f$ and $U_{0}$  for a fixed $\tau=2.0$. In the region that marked red,
the model works as a heat engine while in the region that marked in
green the model acts as a refrigerator. }
\end{figure}

The performance $P_{ref}$ of the refrigerator is explored as a function of model
parameters. At quasistatic limit (regardless of the choice of $A$),
$P_{ref}$ always approaches to the Carnot refrigerator,
\begin{equation}
P_{ref}={1\over \tau-1}.
\end{equation}
In Fig. 7 we have shown $P_{ref}$ as a function of $U_0$. Clearly the figure indicates that the 
performance gets weaker when the medium friction is temperature dependent.
The two operation regions of the engine in parameter space of $f$ and $U_{0}$
is also shown in Fig. 8. In this figure the region marked by red is the region
where the model works as a heat engine while green is the region where the
model acts as a refrigerator.
% The region marked by white is the region where
% the model acts neither as a heat engine nor as refrigerator.

We finally examine the thermodynamic property of the engine by
including the heat exchange via the kinetic and the potential energies.
When the system works as a heat engine, the net flux of the particle is
from hot to the cold heat baths. Similarly, when the engine acts as a
refrigerator, the net flow of the  particle is from the cold to the
hot reservoir. Hence when a particle moves from hot to cold heat bath,
in one cycle, an amount of ${1\over 2}k_{B}(T_h - T_c)$ energy is
transferred via kinetic energy.
\begin{figure}[ht]
\centering
{
\includegraphics[width=8cm]{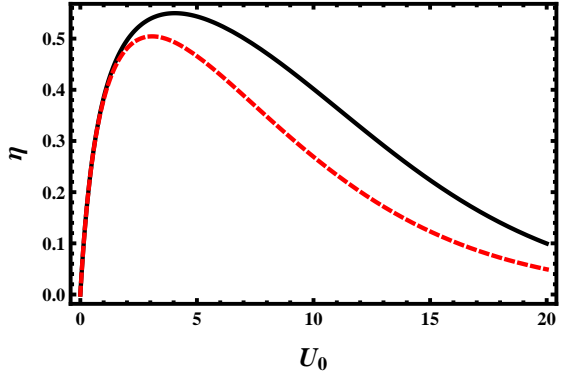}}
\caption{ The efficiency $\eta$ as a function of  $U_{0}$ for the  parameter
values of $f=0.0$, and
$\tau=4.0$. The efficiency is plotted by considering the heat exchange
via kinetic energy.}
\end{figure}

\begin{figure}[ht]
\centering
%\subfigure[Bild a.] % caption for subfigure a
{
    %\label{fig:sub:a}
    \includegraphics[width=8cm]{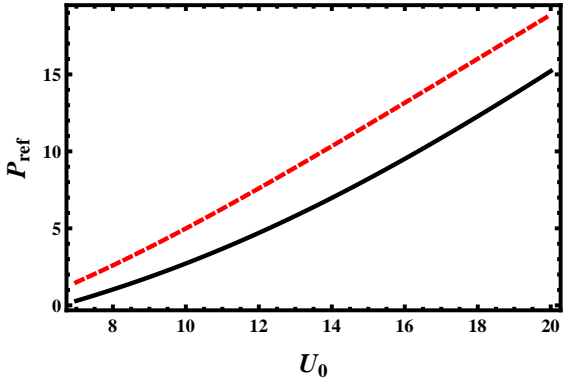}}
\caption{ $P_{ref}$ as a function of $U_0$ for the  parameter values of $
f=0.8$, and   $\tau=6.0$. The solid line stands the current which is
plotted  by taking  $\gamma$ to be temperature dependent while the dashed
line is plotted by taking $\gamma$ to be constant.  The energy exchange
via kinetic energy is taken into account}.
\end{figure}
\begin{figure}[ht]
\centering
{
\includegraphics[width=8cm]{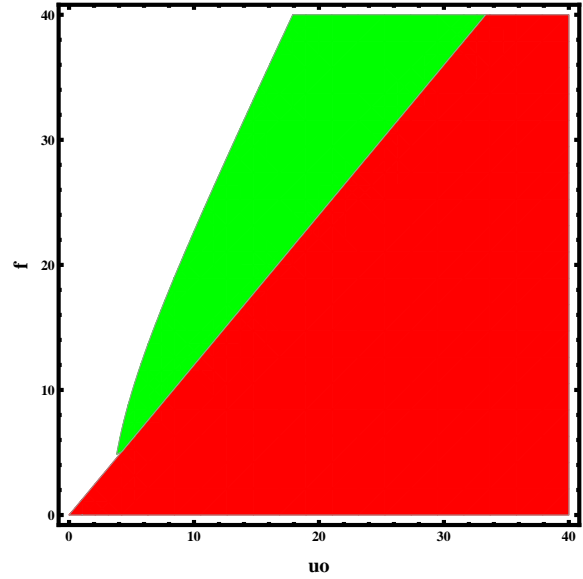}}
\caption{ The phase diagram in parameter space of $U_{0}$ and $f$. In the
region that marked red, the model works as a heat engine while in the region
that marked in green the model acts as a refrigerator. On the other hand,
in the white region, the model acts neither as a heat engine nor as
refrigerator.
}
\end{figure}
When the heat exchange via the kinetic energy is included, Carnot efficiency
will not be attained for the model even at the quasistatic limit. This is due
to the fact that the heat flow via kinetic energy is irreversible. 

We have also examined the dependence of  the efficiency in the presence of the
kinetic energy heat flow for the model parameters. In the quasistatic limit,
the steady state efficiency takes a form
\begin{eqnarray}
\eta^{irr}&=& {\tau-1\over \tau}\Omega
\end{eqnarray}
where
\begin{eqnarray}
\Omega={{4U_0}\over 4U_0 \tau- \tau^2-1}
\end{eqnarray}
Here  $0<\Omega<1$ reveal that the efficiency can
never approaches the Carnot efficiency even at quasistatic.
Hence $\eta^{irr}<\eta$. At the
quasistatic limit, the efficiency remains close to zero for both large and
small values of the barrier height $U_ {0}$ as shown in Fig. 9. However, the
efficiency has an optimal value at certain barrier height.

The heat flow via the kinetic energy has also influence on the performance
of the refrigerator. Our analytical result show that the coefficient of performance
of the refrigerator is always less than the Carnot refrigerator when the engine
operates quasistatically. In the quasistatic limit, the steady state
$P_{irr}^{ref}$ converges to
\begin{eqnarray}
P_{irr}^{ref}(t)= {1 \over \tau-1}\Psi
\end{eqnarray}
where
\begin{eqnarray}
\Psi={{{{4U_0}\over\tau}-\tau^2+1}\over {4U_0}}.
\end{eqnarray}
Again here  $0<\Psi<1$
reveals that Carnot refrigerator is unattainable
even at quasistatic limit. When the engine operates at
finite time, the system exhibits a higher performance of the refrigerator  for constant $\gamma$ case (see Fig. 10).

A complete picture for the operation regions of the heat engine is obtained
by observing the phase diagram in parameter space of $U_{0}$ and $f$ as 
shown in Fig.11. Again here we use the same color scheme where the red shows
the region where the model works as a heat engine while green represents the
region for a refrigerator. In the region that marked white, the model acts
neither as a heat engine nor as refrigerator

\section {Summary and conclusion}
In this work, we study the effect of temperature on the performance
of the heat engine as well as on its mobility by considering a viscous
friction that has an exponential temperature dependence. Our analysis
shows that whether the viscous friction is temperature dependent or
not,  at quasistatic limit,  one always gets Carnot efficiency and Carnot
refrigerator provided that the heat exchange via kinetic energy is neglected.
However, when the heat exchange via the kinetic energy is included, both Carnot
efficiency and Carnot refrigerator are unattainable even at quasistatic
limit. Meanwhile, far from quasistatic limit, the engine exhibits an
enhanced performance when the viscose friction is taken to be
temperature dependent. Our detail analysis indicates that the thermal
background temperature has dual effects as it weakens the strength of the
viscous friction.

In conclusion, in this work, we present a pragmatic model system that
not only serves as a basic understanding of non-equilibrium  physics
but also for construction of artificial tiny motors that operate at
microscopic or nanoscopic levels. Our study depicts that the role
of temperature is twofold. It enhances the performance
of the motor directly by assisting the particle to surmount  the
potential barrier or indirectly by weakening the intensity of the
viscous friction.

{\it Acknowledgment.\textemdash} 
 We would like to thank Mulugeta Bekele for the interesting discussions we had.

\end{document}